\documentclass[pre,aps,showkeys,showpacs,superscriptaddress,twocolumn,%
amsmath,amssymb]{revtex4} \usepackage{dsfont,graphicx}

\newcommand{\gl}[1]{\eqref{#1}}
\newcommand{\la}{\left\langle}
\newcommand{\ra}{\right\rangle}
\newcommand{\al}{\alpha}
\newcommand{\be}{\beta}

\newcommand{\df}{\Delta f}
\newcommand{\dfhat}{\widehat{\df}}

\newcommand{\pf}{p_{0}}
\newcommand{\pr}{p_{1}}
\newcommand{\pa}{p_{\al}}
\newcommand{\Ua}{U_{\al}}
\newcommand{\Uhat}{\widehat{U}}
\newcommand{\Uhatvar}{\Uhat^{\scriptscriptstyle(I\!I)}}
\newcommand{\Xhat}{\widehat{X}}

\newcommand{\Var}{\operatorname{Var}}
\newcommand{\widebar}[1]{\overline{#1}}

\begin{document}

\title{Measuring the convergence of Monte Carlo free energy calculations}
\author{Aljoscha M. Hahn} \altaffiliation{Present address: Technische
  Universit\"at Berlin, Institut f\"ur Theo\-re\-ti\-sche Physik,
  10623 Berlin, Germany} \affiliation{Institut f\"ur Physik, Carl von
  Ossietzky Universit\"at, 26111 Oldenburg, Germany} \author{Holger Then}
\altaffiliation{Present address: University of Bristol, Department of
  Mathematics, University Walk, Bristol BS8 1TW, UK}
\affiliation{Institut f\"ur Physik, Carl von Ossietzky Universit\"at,
  26111 Oldenburg, Germany}

\begin{abstract}
  The nonequilibrium work fluctuation theorem provides the way for calculations
  of (equilibrium) free energy based on work measurements of nonequilibrium,
  finite-time processes and their reversed counterparts by applying Bennett's
  acceptance ratio method. A nice property of this method is that each free
  energy estimate readily yields an estimate of the asymptotic mean square
  error. Assuming convergence, it is easy to specify the uncertainty of the
  results. However, sample sizes have often to be balanced with respect to
  experimental or computational limitations and the question arises whether
  available samples of work values are sufficiently large in order to ensure
  convergence. Here, we propose a convergence measure for the two-sided free
  energy estimator and characterize some of its properties, explain how it
  works, and test its statistical behavior. In total, we derive a convergence
  criterion for Bennett's acceptance ratio method.
\end{abstract}


\maketitle

\section{Introduction}\label{sec:1}

Many methods have been developed in order to estimate free energy differences,
ranging from thermodynamic integration \cite{Kirkwood1935,Gelman1998}, path
sampling \cite{Minh2009}, free energy perturbation \cite{Zwanzig1954}, umbrella
sampling \cite{Torrie1977,Chen1997,Oberhofer2008}, adiabatic switching
\cite{Watanabe1990}, dynamic methods
\cite{Sun2003,Ytreberg2004,Jarzynski2006,Ahlers2008}, optimal protocols
\cite{Then2008,Geiger2010}, asymptotic tails \cite{vonEgan-Krieger2008}, to
targeted and escorted free energy perturbation
\cite{Meng2002,Jarzynski2002,Oberhofer2007,Vaikuntanathan2008,Hahn2009a}.
Yet, the reliability and efficiency of the approaches have not been considered
in full depth. Fundamental questions remain unanswered \cite{Lu2007}, e.g.,
what method is best for evaluating the free energy? Is the free energy estimate
reliable and what is the error in it? How can one assess the quality of the
free energy result when the true answer is unknown? Generically, free energy
estimators are strongly biased for finite sample sizes, such that the bias
constitutes the main source of error of the estimates. Moreover, the bias can
manifest itself in a seemingly convergence of the calculation by reaching a
stable value, although far apart from the desired true value. Therefore, it is
of considerable interest to have reliable criteria for the convergence of free
energy calculations.

Here we focus on the convergence of Bennett's acceptance ratio method. Thereby,
we will only be concerned with the intrinsic statistical errors of the method
and assume uncorrelated and unbiased samples from the work densities. For
incorporation of instrument noise, see Ref. \cite{Maragakis2008}.

With emerging results from nonequilibrium stochastic thermodynamics,
Bennett's acceptance ratio method
\cite{Bennett1976,Meng1996,Kong2003,Shirts2008} has revived actual interest.

Recent research has shown that the isothermal free energy difference
$\df=f_1-f_0$ of two thermal equilibrium states $0$ and $1$, both at the same
temperature $T$, can be determined by externally driven nonequilibrium
processes connecting these two states. In particular, if we start the process
with the initial thermal equilibrium state $0$ and perturb it towards $1$ by
varying the control parameter according to a predefined protocol, the work $w$
applied to the system will be a fluctuating random variable distributed
according to a probability density $\pf(w)$. This direction will be denoted
with \textit{forward}. Reversing the process by starting with the initial
equilibrium state $1$ and perturbing the system towards $0$ by the time
reversed protocol, the work $w$ done $by$ the system in the \textit{reverse}
process will be distributed according to a density $\pr(w)$. Under some quite
general conditions, the forward and reverse work densities $\pf(w)$ and
$\pr(w)$ are related to each other by Crooks fluctuation theorem
\cite{Crooks1999,Campisi2009}
\begin{align}\label{fth}
  \frac{\pf(w)}{\pr(w)}=e^{w-\df}.
\end{align}
Throughout the paper, all energies are understood to be measured in units of
the thermal energy $kT$, where $k$ is Boltzmann's constant. The fluctuation
theorem relates the equilibrium free energy difference $\df$ to the
nonequilibrium work fluctuations which permits calculation (estimation) of
$\df$ using samples of work-values measured either in only one direction
(\textit{one-sided} estimation) or in both directions (\textit{two-sided}
estimation). The one-sided estimators rely on the Jarzynski relation
\cite{Jarzynski1997} $e^{-\df}=\int e^{-w} \pf(w) dw$ which is a direct
consequence of Eq.~\gl{fth}, and the free energy is estimated by calculating
the sample mean of the exponential work. In general, however, it is of great
advantage to employ optimal two-sided estimation with Bennett's acceptance
ratio method \cite{Bennett1976}, although one has to measure work-values in
both directions.

The work fluctuations necessarily allow for events which ``violate'' the second
law of thermodynamics such that $w<\df$ holds in forward direction and $w>\df$
in reverse direction, and the accuracy of any free energy estimate solely based
on knowledge of Eq.~\gl{fth} will strongly depend on the extend to which these
events are observed. The fluctuation theorem indicates that such events will in
general be exponentially rare; at least, it yields the inequality
$\la w \ra_1 \leq \df \leq \la w \ra_0$ \cite{Jarzynski1997}, which states the
second law in terms of the average work $\la w\ra_0$ and $\la w\ra_1$ in
forward and reverse direction, respectively. Reliable free energy calculations
will become harder the larger the dissipated work $\la w \ra_0 - \df$ and
$\df - \la w \ra_1$ in the two directions is \cite{Hahn2009a}, i.e.\ the
farther from equilibrium the process is carried out, resulting in an increasing
number $N$ of work values needed for a converging estimate of $\df$. This
difficulty can also be expressed in terms of the overlap area
$\mathcal{A}=\int\min\{\pf(w),\pr(w)\}dw \leq 1$ of the work densities, which
is just the sum of the probabilities $\int_{-\infty}^{\df}p_0 dw$ and
$\int_{\df}^{\infty}p_1 dw$  of observing second-law ``violating'' events in the
two directions. Hence, $N$ has to be larger than $1/\mathcal{A}$. However, an
\textit{\`a priori} determination of the number $N$ of work values required
will be impossible in situations of practical interest. Instead, it may be
possible to determine \textit{\`a posteriori} whether a given calculation of
$\df$ has converged. The present paper develops a criterion for the convergence
of two-sided estimation which relies on monitoring the value of a suitably
bounded quantity $a$, the convergence measure. As a key feature, the
convergence measure $a$ checks if the relevant second-law ``violating'' events
are observed sufficiently and in the right proportion for obtaining an accurate
and precise estimate of $\df$.

Two-sided free energy estimation, i.e.\ Bennett's acceptance ratio method,
incorporates a pair of samples of both directions: given a sample $\{w^0_k\}$
of $n_0$ forward work values, drawn independently from $\pf(w)$, together with
a sample $\{w^1_l\}$ of $n_1$ reverse work values drawn from $\pr(w)$, the
asymptotically optimal estimate $\dfhat$ of the free energy difference $\df$ is
the unique solution of \cite{Bennett1976,Meng1996,Kong2003,Shirts2008}
\begin{align}\label{benest}
  \frac{1}{n_0} \sum\limits_{k=1}^{n_0} \frac{1}{\be + \al e^{w^0_k-\dfhat}}
  = \frac{1}{n_1} \sum\limits_{l=1}^{n_1} \frac{1}{\al + \be e^{-w^1_l+\dfhat}},
\end{align}
where $\al$ and $\be\in(0,1)$ are the fraction of forward and reverse work
values used, respectively,
\begin{align}
  \al=\frac{n_0}{N} \quad \text{and} \quad \be=\frac{n_1}{N},
\end{align}
with the total sample size $N=n_0+n_1$.

Originally found by Bennett \cite{Bennett1976} in the context of free energy
perturbation \cite{Zwanzig1954}, with ``work'' being simply an energy
difference, the two-sided estimator \gl{benest} was generalized by Crooks
\cite{Crooks2000} to actual work of nonequilibrium finite time processes. We
note that the two-sided estimator has remarkably good properties
\cite{Bennett1976,Meng1996,Shirts2003,Hahn2009a}. Although in general biased
for small sample sizes $N$, the bias
\begin{align}\label{bias}
  b = \la \dfhat - \df \ra
\end{align}
asymptotically vanishes for $N\to\infty$, and the estimator is the one with
least mean square error (viz.\ variance) in the limit of large sample sizes
$n_0$ and $n_1$ within a wide class of estimators. In fact, it is the optimal
estimator if no further knowledge on the work densities besides the fluctuation
theorem is given \cite{Hahn2009a,Maragakis2008}. It comprises one-sided
Jarzynski estimators as limiting cases for $\al\to0$ and $\al\to1$,
respectively. Recently \cite{Hahn2009b}, the asymptotic mean square error has
been shown to be a convex function of $\al$ for fixed $N$, indicating that
typically two-sided estimation is superior if compared to one-sided estimation.

In the limit of large $N$, the mean square error
\begin{align}\label{msefull}
  m = \la(\dfhat-\df)^2\ra
\end{align}
converges to its asymptotics
\begin{align}\label{mse}
  X(N,\al) = \frac{1}{N} \frac{1}{\al\be} \big( \frac{1}{\Ua}-1 \big),
\end{align}
where the overlap (integral) $\Ua$ is given by
\begin{align}\label{Udef}
  \Ua = \int\limits \frac{\pf\pr}{\al\pf+\be\pr} dw.
\end{align}
Likewise, in the large $N$ limit the probability density of the estimates
$\dfhat$ (for fixed $N$ and $\al$) converges to a Gaussian density with mean
$\df$ and variance $X(N,\al)$ \cite{Meng1996}. Thus, within this regime a
reliable confidence interval for a particular estimate $\dfhat$ is obtained
with an estimate $\Xhat(N,\al)$ of the variance,
\begin{align}\label{Xhat}
  \Xhat(N,\al) := \frac{1}{N\al\be} \big( \frac{1}{\Uhat_\al}-1 \big),
\end{align}
where the overlap estimate $\Uhat_\al$ is given through
\begin{align}\label{Uhat}
  \Uhat_\al
  := \frac{1}{n_0} \sum\limits_{k=1}^{n_0} \frac{1}{\be + \al e^{w^0_k-\dfhat}}
  = \frac{1}{n_1} \sum\limits_{l=1}^{n_1} \frac{1}{\al + \be e^{-w^1_l+\dfhat}}.
\end{align}

To get some feeling for when the large $N$ limit ``begins'', we state a close
connection between the asymptotic mean square error and the overlap area
$\mathcal{A}$ of the work densities as follows:
\begin{align}\label{mseineq}
  \frac{1-2\mathcal{A}}{N\mathcal{A}} < X(N,\al) \leq
  \frac{1-\mathcal{A}}{\al\be N\mathcal{A}},
\end{align}
see Appendix \ref{appendix.A}. Using $\al\approx0.5$ and assuming that the
estimator has converged once $X<1$, we find the ``onset'' of the large $N$
limit for $N>\frac{1}{\mathcal{A}}$. However, this onset may actually be one or
more orders of magnitude larger.

\begin{figure}
  \includegraphics{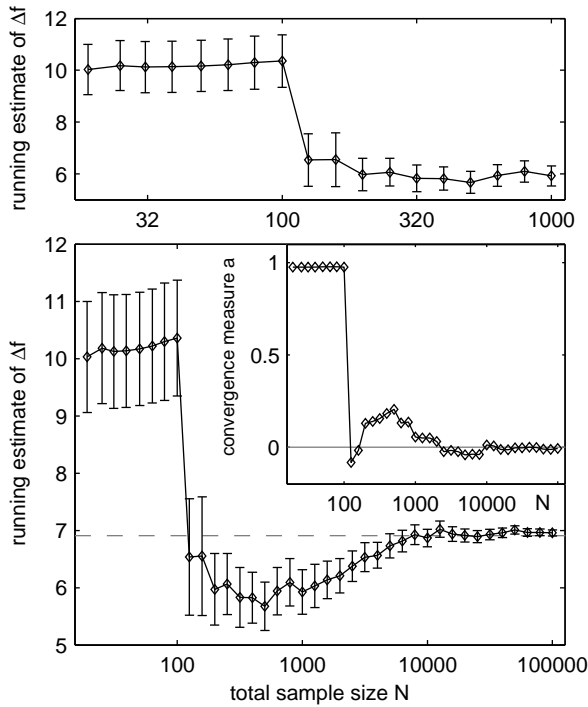}
  \caption{\label{fig:1}Displayed are free energy estimates $\dfhat$ in
    dependence of the sample size $N$, reaching a seemingly stable plateau if
    $N$ is restricted to $N=1000$ (top panel). Another stable plateau is
    reached if the sample size is increased up to $N=100\,000$ (bottom panel).
    Has the estimate finally converged? The answer is given by the
    corresponding graph of the convergence measure $a$ which is shown in the
    inset. The fluctuations around zero indicate convergence. The exact value
    of the free energy difference is visualized by the dashed horizontal line.}
\end{figure}

If we do not know whether the large $N$ limit is reached, we cannot state a
reliable confidence interval of the free energy estimate: a problem which
encounters frequently within free energy calculations is that the estimates
``converge'' towards a stable plateau. While the sample variance can become
small, it remains unclear whether the reached plateau represents the correct
value of $\df$. Possibly, the found plateau is subject to some large bias,
i.e.\ far off the correct value. A typical situation is displayed in
Fig.~\ref{fig:1} which shows successive two-sided free energy estimates in
dependence of the sample size $N$. The errorbars are obtained with an
error-propagation formula for the variance of $\dfhat$ which reflects the
sample variances, see Appendix \ref{appendix.C} after reading Sec.~\ref{sec:3}.
If we take a look on the top panel of Fig.~\ref{fig:1}, we might have the
impression that the free energy estimate has converged at $N\approx300$
already, while the bottom panel reaches out to larger sample sizes where it
becomes visible that the ``convergence'' in the top panel was just pretended.
Finally, we may ask if the estimates shown in the bottom panel have converged
at $N\gtrsim10000$? As we know the true value of $\df$, which is depicted in
the figure as a dashed line, we can conclude that convergence actually
happened.

The main result of the present paper is the statement of a convergence
criterion for two-sided free energy estimation in terms of the
\textit{behavior} of the convergence measure $a$. As will be seen, $a$
converges to zero. Moreover, this happens almost simultaneously with the
convergence of $\dfhat$ to $\df$. The procedure is as follows: While drawing an
increasing number of work values in both directions (with fixed fraction $\al$
of forward draws), successive estimates $\dfhat$ and corresponding values of
$a$, based on the present samples of work, are calculated. The values of $a$
are displayed graphically in dependence of $N$, preferably on a log-scale. Then
the typical situation observed is that $a$ is close to it's upper bound for
small sample sizes $N<\frac{1}{\mathcal{A}}$, which indicates lack of ``rare
events'' which are required in the averages of Eq.~\gl{benest} (i.e.\ those
events which ``violate'' the second law). Once $N$ becomes comparable to
$\frac{1}{\mathcal{A}}$, single observations of rare events happen and change
the value of $\dfhat$ and $a$ rapidly. In this regime of $N$, rare events are
likely to be observed either disproportionally often or seldom, resulting in
strong fluctuations of $a$ around zero. This indicates the transition region to
the large $N$ limit. Finally, for some $N\gg\frac{1}{\mathcal{A}}$, the large $N$
limit is reached, and $a$ typically fluctuates close around zero, cf.\ the
inset of Fig.~\ref{fig:1}.

The paper is organized as follows. In Sec.~\ref{sec:2}, we first consider a
simple model for the source of bias of two-sided estimation which is intended to
obtain some insight into the convergence properties of two-sided estimation.
The convergence measure $a$, which is introduced in Sec.~\ref{sec:3}, however,
will not depend on this specific model. As the convergence measure is based on
a sample of forward and reverse work values, it is itself a random variable,
raising the question of reliability once again. Using numerically simulated
data, the statistical properties of the convergence measure will be elaborated
in Sec.~\ref{sec:4}. The convergence criterion is stated in Sec.~\ref{sec:5},
and Sec.~\ref{sec:6} presents an application to the estimation of the chemical
potential of a Lennard-Jones fluid.

\section{Neglected tail model for two-sided estimation}\label{sec:2}

To obtain some first qualitative insight into the relation between the
convergence of Eq.~\gl{Uhat} and the bias of the estimated free energy
difference, we adopt the neglected tails model \cite{Wu2004} originally
developed for one-sided free energy estimation.

Two-sided estimation of $\df$ essentially means estimating the overlap $\Ua$
from two sides, however in a dependent manner, as $\dfhat$ is adjusted such
that both estimates are equal in Eq.~\gl{Uhat}.

\begin{figure}
  \includegraphics{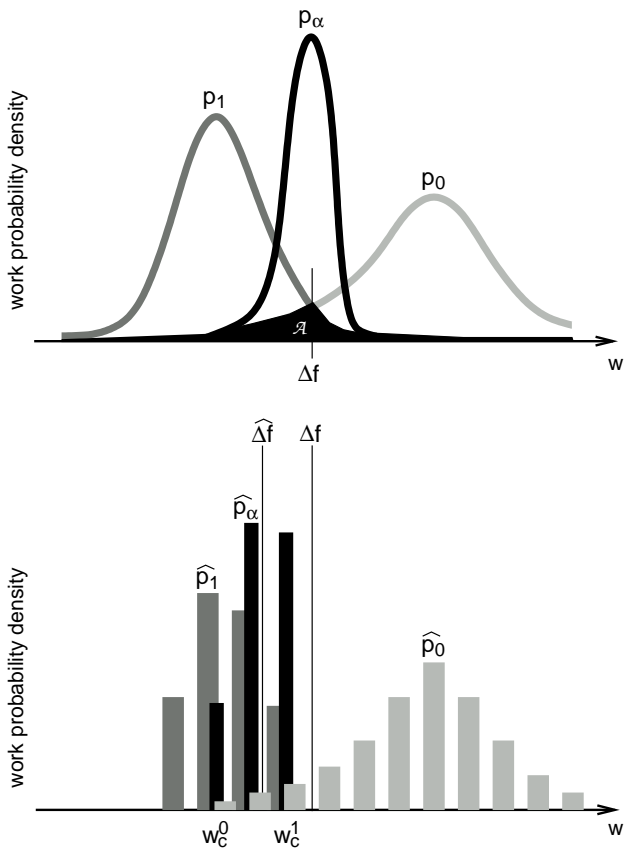}
  \caption{\label{fig:2}Schematic diagram of reverse $\pr$, overlap $\pa$, and
    forward $\pf$ work densities (top). Schematic histograms of finite samples
    from $\pf$ and $\pr$, where in particular the latter is imperfectly
    sampled, resulting in a biased estimate $\dfhat$ of the free energy
    difference (bottom).}
\end{figure}

Consider the (normalized) overlap density $\pa(w)$, defined as harmonic mean of
$\pf$ and $\pr$:
\begin{align}\label{pa}
  \pa(w) = \frac{1}{\Ua} \frac{\pf(w)\pr(w)}{\al\pf(w)+\be\pr(w)}.
\end{align}
For $\al\to0$ and $\al\to1$, $\pa$ converges to $\pf$ and $\pr$, respectively.
The dominant contributions to $\Ua$ come from the overlap region of $\pf$ and
$\pr$ where $\pa$ has its main probability mass, see Fig.~\ref{fig:2} (top).

In order to obtain an accurate estimate of $\df$ with the two-sided estimator
\gl{benest}, the sample $\{w^0_k\}$ drawn from $\pf$ has to be representative
for $\pf$ up to the \textit{overlap region} in the left tail of $\pf$, and the
sample $\{w^1_k\}$ drawn from $\pr$ has to be representative for $\pr$ up to
the overlap region in the right tail of $\pr$. For small $n_0$ and $n_1$,
however, we will have certain effective cut-off values $w^0_c$ and $w^1_c$ for
the samples from $\pf$ and $\pr$, respectively, beyond which we typically will
not find any work values, see Fig.~\ref{fig:2} (bottom).

We introduce a model for the bias \gl{bias} of two-sided free energy estimation
as follows. Assuming a ``semi-large'' $N=n_0+n_1$, the \textit{effective}
behavior of the estimator for fixed $n_0$ and $n_1$ is modeled by substituting
the sample averages appearing in the estimator \gl{benest} with ensemble
averages, however truncated at $w^0_c$ and $w^1_c$, respectively:
\begin{align}\label{negtail}
  \int\limits_{w^0_c}^{\infty} \frac{\pf(w)}{\be + \al e^{w-\la\dfhat\ra}} dw
  = \int\limits_{-\infty}^{w^1_c}\frac{\pr(w)}{\al + \be e^{-w+\la\dfhat\ra}} dw .
\end{align}
Thereby, the cut-off values $w^i_c$ are thought fixed (only depending on $n_0$
and $n_1$) and the expectation $\la \dfhat \ra$ is understood to be the unique
root of Eq.~\gl{negtail}, thus being a function of the cut-off values $w^i_c$,
$i=0,1$.

In order to elaborate the implications of this model, we rewrite
Eq.~\gl{negtail} with the use of the fluctuation theorem \gl{fth} such that the
integrands are equal,
\begin{align}\label{negtail2}
  e^{\la\dfhat-\df\ra} = \frac{\int\limits_{-\infty}^{w^1_c}
    \frac{\pf(w)}{\al e^{w-\la\dfhat\ra} + \be} dw}{
    \int\limits_{w^0_c}^{\infty} \frac{\pf(w)}{\al e^{w-\la\dfhat\ra} + \be} dw},
\end{align}
and consider two special cases:
\begin{enumerate}
\item \textit{Large $n_1$ limit}: Assume the sample size $n_1$ is large enough
  to ensure that the overlap region is fully and accurately sampled (large
  $n_1$ limit). Thus, $w^1_c$ can be safely set equal to $\infty$ in
  Eq.~\gl{negtail2}, and the r.h.s.\ becomes larger than unity. Accordingly,
  our model predicts a positive bias.
\item \textit{Large $n_0$ limit}: Turning the tables and using $w^0_c=-\infty$
  in Eq.~\gl{negtail2}, the model implies a negative bias.
\end{enumerate}

In essence, $\la\dfhat\ra$ is shifted away from $\df$ towards the
insufficiently sampled density. In general, when none of the densities is
sampled sufficiently, the bias will be a trade off between the two cases.

Qualitatively, from the neglected tails model, we find the main source of bias
resulting from a different convergence behavior of forward and reverse
estimates \gl{Uhat} of $\Ua$. The task of the next section will be to develop a
quantitative measure of convergence.

\section{The convergence measure}\label{sec:3}

In order to check convergence, we propose a measure which relies on a
consistency check of estimates based on first and second moments of the Fermi
functions that appear in the two-sided estimator \gl{Uhat}. In a recent study
\cite{Hahn2009a}, we already used this measure for the special case of
$\al=\frac{1}{2}$. Here, we give a generalization to arbitrary $\al$, study the
convergence measure in greater detail, and justify its validity and usefulness.
In the following we will assume that the densities $p_0$ and $p_1$ have the
same support.

It was discussed in the preceding section that the large $N$ limit is reached
and hence the bias of two-sided estimation vanishes if the overlap $\Ua$ is (in
average) correctly estimated from both sides, $0$ and $1$. Defining the
complementary Fermi functions $t_c(w)$ and $b_c(w)$ (for given $\al$) with
\begin{align}
 &t_c(w) = \frac{1}{\al + \be e^{-w+c}}, \nonumber \\
 &b_c(w) = \frac{1}{\al e^{w-c} + \be}, \label{fermidef}
\end{align}
such that $\al t_c(w)+\be b_c(w)=1$ and $t_c(w)=e^{w-c}b_c(w)$ holds. The
overlap \gl{Udef} can be expressed in terms of first moments,
\begin{align}\label{Udef2}
 \Ua = \int t_{\scriptscriptstyle\df}(w) p_1(w) dw
 = \int b_{\scriptscriptstyle\df}(w) p_0(w) dw,
\end{align}
and the overlap estimate $\Uhat_\al$, Eq.~\gl{Uhat}, is simply obtained by
replacing in Eq.~\gl{Udef2} the ensemble averages by sample averages,
\begin{align}\label{Uhatnew}
 \Uhat_\al = \widebar{t_{\scriptscriptstyle\dfhat}}^{\scriptscriptstyle (1)}
 = \widebar{ b_{\scriptscriptstyle\dfhat}}^{\scriptscriptstyle(0)}.
\end{align}
According to Eq.~\gl{benest}, the value of $\dfhat$ is defined such that the
above relation holds. Note that $\dfhat = \dfhat(w^0_1,\ldots,w^1_{n_1})$ is a
single-valued function depending on all work values used in both directions.
The overbar with index $(i)$ denotes an average with a sample $\{w^i_k\}$ drawn
from $p_i$, $i=0,1$. For an arbitrary function $g(w)$ it explicitly reads
\begin{align}\label{sampleaverage}
 \widebar{g}^{\scriptscriptstyle(i)}
 = \frac{1}{n_i} \sum\limits_{k=1}^{n_i} g(w^i_k)
\end{align}
Interestingly, $\Ua$ can be expressed in terms of second moments of the Fermi
functions such that it reads
\begin{align}\label{Udef3}
 \Ua = \al \int t_{\scriptscriptstyle\df}^2 p_1 dw
 + \be \int b_{\scriptscriptstyle\df}^2 p_0 dw
\end{align}
A useful test of self-consistency is to compare the first order estimate $\Uhat_\al$, with the
second order estimate $\Uhatvar_\al$, where the latter is defined by replacing the ensemble
averages in Eq.~\gl{Udef3} with sample averages:
\begin{align}\label{Uhat2}
 \Uhatvar_\al
 = \al \widebar{t_{\scriptscriptstyle\dfhat}^2}^{\scriptscriptstyle (1)}
 + \be \widebar{ b_{\scriptscriptstyle\dfhat}^2}^{\scriptscriptstyle(0)}.
\end{align}
Thereby, the estimates $\dfhat$, $\Uhat_\al$, and $\Uhatvar_\al$, are
understood to be calculated with the same pair of samples $\{w^0_k\}$ and
$\{w^1_l\}$.

The relative difference of this comparison results in the definition of the
convergence measure,
\begin{align}\label{a}
  a = \frac{\Uhat_\al-\Uhatvar_\al}{\Uhat_\al},
\end{align}
for all $\al\in(0,1)$. Clearly, in the large $N$ limit, $a$ will converge to
zero, as then $\dfhat$ converges to $\df$ and thus $\Uhat_\al$ as well as
$\Uhatvar_\al$ converge to $\Ua$. As argued below, it is the estimate
$\Uhatvar_\al$ that converges last, hence $a$ converges somewhat later than
$\dfhat$.

Below the large $N$ limit, $a$ will deviate from zero. From the general
inequality
\begin{align}\label{Uineq}
  \Uhat_\al^2 \le \Uhatvar_\al < 2 \Uhat_\al
\end{align}
(see Appendix \ref{appendix.B}) follow upper and lower bounds on $a$ which read
\begin{align}\label{aineq1}
  -1 < a \le 1-\Uhat_\al < 1.
\end{align}
The behavior of $a$ with increasing sample size $N=n_0+n_1$ (while keeping the
fraction $\al=\frac{n_0}{N}$ constant) can roughly be characterized as follows:
$a$ ``starts'' close to its upper bound for small $N$ and decreases towards
zero with increasing $N$. Finally, $a$ begins to fluctuate around zero when the
large $N$ limit is reached, i.e.\ when the estimate $\dfhat$ converges.

\begin{figure}
  \includegraphics{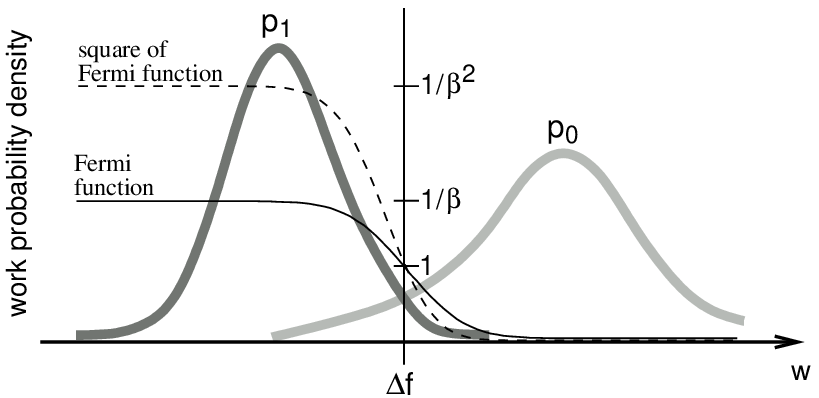}
  \caption{\label{fig:3}Schematic plot which shows that the forward work
    density, $\pf(w)$, samples the Fermi function
    $b_{\scriptscriptstyle\df}(w)=1/(\be+\al e^{w-\df})$ somewhat earlier than its
    square.}
\end{figure}

To see this qualitatively, we state that the second order estimate
$\Uhatvar_\al$ converges later than the first order estimate $\Uhat_\al$, as
the former requires sampling the tails of $\pf$ and $\pr$ to a somewhat wider
extend than the latter, cf.\ Fig.~\ref{fig:3}. For small $N$, both, $\Uhat_\al$
and $\Uhatvar_\al$, will typically underestimate $\Ua$, as the ``rare-events''
which contribute substantially to the averages \gl{Uhatnew} and \gl{Uhat2} are
quite likely not to be observed sufficiently, if at all. For the same reason,
generically $\Uhatvar_\al < \Uhat_\al$ will hold, since
$b_{\scriptscriptstyle\dfhat}(w^0)^2\leq b_{\scriptscriptstyle\dfhat}(w^0)$
holds for $w^0\geq\dfhat$ and similar
$t_{\scriptscriptstyle\dfhat}(w^1)^2\leq t_{\scriptscriptstyle\dfhat}(w^1)$ for
$w^1\leq\dfhat$. Therefore, $a$ is typically positive for small $N$. In
particular, if $N$ is so small that \textit{all} work values of the forward
sample are larger than $\dfhat$ and all work values of the reverse sample are
smaller than $\dfhat$, then $\Uhatvar_\al$ becomes much smaller than
$\Uhat_\al$, resulting in $a\approx 1$.

Analytic insight into the behavior of $a$ for small $N$ results from the fact
that $n \widebar{x}^2 \geq \widebar{x^2}$ for any set $\{x_1,\ldots x_n\}$ of
positive numbers $x_k$. Using this in Eq.~\gl{Uhat2} yields
\begin{align}\label{Uineq2}
  \Uhatvar_\al \leq 2N\al\be\Uhat_\al^2
\end{align}
and
\begin{align}\label{aineq2}
  1 - 2\al\be N\Uhat_\al \le a \le 1 - \Uhat_\al.
\end{align}
This shows that as long as $N\Uhat_\al \ll 1$ holds, $a$ is close to its upper
bound $1-\Uhat_\al \approx 1$. In particular, if $\al=\frac{1}{2}$ and $N=2$,
then $a=1-\Uhat_\al$ holds exactly.

Averaging the inequality for some $N$ sufficiently large to ensure
$\la a \ra \approx 0$ and $\la \Uhat_\al \ra \approx \Ua$, we get a lower bound
on $N$ which reads $N\geq \frac{1}{2\al\be\Ua}$. Again, this bound can be
related to the overlap area $\mathcal{A}$: taking $\al=\frac{1}{2}$ and using
$U_{\frac{1}{2}}\leq 2\mathcal{A}$ (see Appendix \ref{appendix.A}), we obtain
$N\geq\frac{1}{\mathcal{A}}$, in concordance with the lower bound for the large
$N$ limit stated in Sec.~\ref{sec:1}.

Last we note that the convergence measure $a$ can also be understood as a
measure of the sensibility of relation \gl{benest} with respect to the value of
$\dfhat$: in the low $N$ regime, the relation is highly sensible to the value
of $\dfhat$, resulting in large values of $a$, whereas in the limit of large
$N$, relation \gl{benest} becomes insensible to small perturbations of
$\dfhat$, corresponding to $a\approx0$. The details are summarized in Appendix
\ref{appendix.D}.

\section{Study of statistical properties of the convergence measure}%
\label{sec:4}

\begin{figure}
  \includegraphics{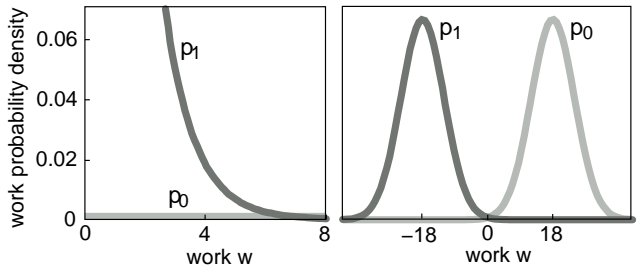}
  \caption{\label{fig:4}Exponential (left panel) and Gaussian (right panel)
    work densities.}
\end{figure}

In order to demonstrate the validity of $a$ as a measure of convergence of
two-sided free energy estimation, we apply it to two qualitatively different
types of work densities, namely exponential and Gaussian, see Fig.~\ref{fig:4}.
Samples from these densities are easily available by standard (pseudo) random
generators. Statistical properties of $a$ are obtained by means of independent
repeated calculations of $\dfhat$ and $a$. While the two types of densities
used are fairly simple, they are entirely different and general enough to
reflect the statistical properties of the convergence measure.

\subsection{Exponential work densities}

The first example uses exponential work densities, i.e.\ 
\begin{align}\label{expdist}
  p_i(w) = \frac{1}{\mu_i} e^{-\frac{w}{\mu_i}}, \quad w \geq 0,
\end{align}
$\mu_i>0$, $i=0,1$. According to the fluctuation theorem \gl{fth}, the mean
values $\mu_i$ of $\pf$ and $\pr$ are related to each other,
$\mu_1=\frac{\mu_0}{1+\mu_0}$, and the free energy difference is known to be
$\df=\ln(1+\mu_0)$.

\begin{figure}
  \includegraphics{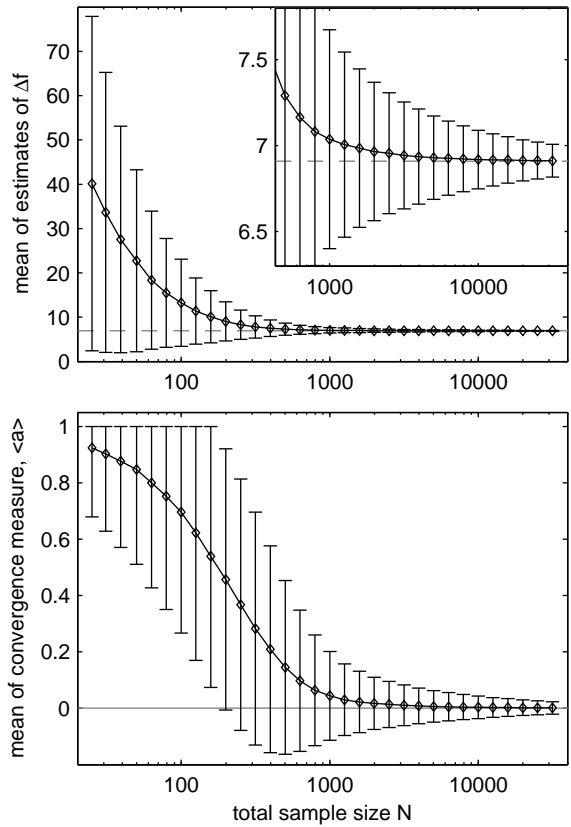}
  \caption{\label{fig:5}Statistics of two-sided free energy estimation
    (exponential work densities): shown are averaged estimates of $\df$ in
    dependence of the total sample size $N$. The errorbars reflect the standard
    deviation. The dashed line shows the exact value of $\df$, and the inset
    the details for large $N$ (top). Statistics of the convergence measure $a$
    corresponding to the estimates of the top panel: shown are the average
    values of $a$ together with their standard deviation in dependence of the
    sample size $N$. Note the characteristic convergence of $a$ towards zero in
    the large $N$ limit (bottom).}
\end{figure}

\begin{figure}
  \includegraphics{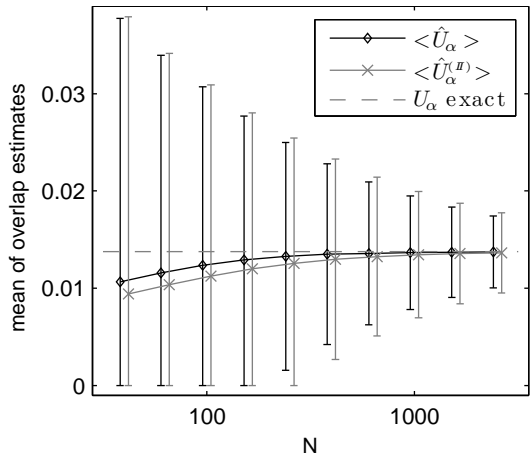}
  \caption{\label{fig:6}Mean values of overlap estimates $\Uhat_\al$ and
    $\Uhatvar_\al$ of first and second order, respectively. The slightly slower
    convergence of $\Uhatvar_\al$ towards $\Ua$ results in the characteristic
    properties of the convergence measure $a$. To enhance clarity, data points
    belonging to the same value of $N$ are spread.}
\end{figure}

\begin{figure}
  \includegraphics{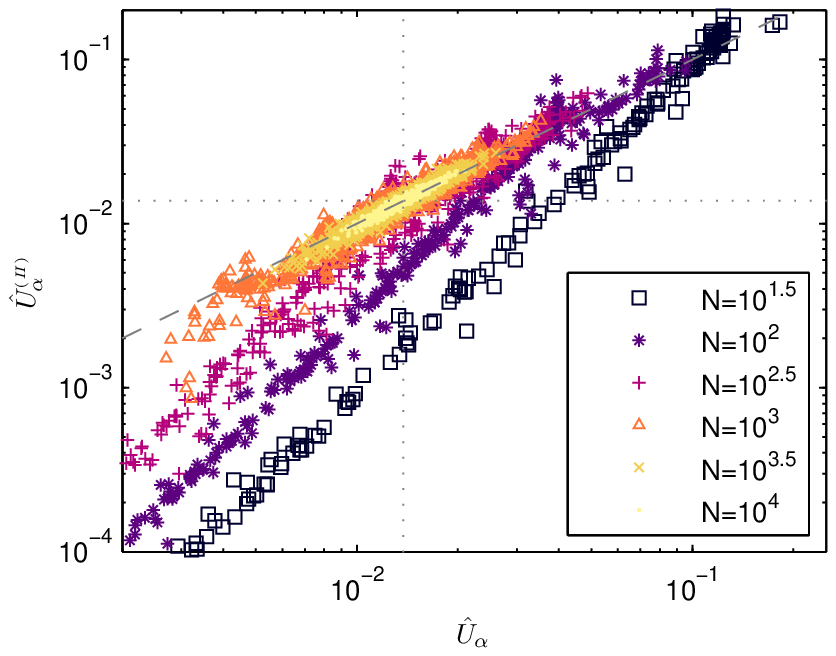}
  \caption{\label{fig:7}(Color online)
    Double-logarithmic scatter plot of
    $\Uhatvar_\al$ versus $\Uhat_\al$ for many individual estimates in
    dependence of the sample size $N$. The dotted lines mark the exact value of
    $\Ua$ on the axes, and the dashed line is the bisectrix
    $\Uhatvar_\al=\Uhat_\al$. The approximatively linear relation between the
    logarithms of $\Uhatvar_\al$ and $\Uhat_\al$ is continued up to the
    smallest observed values ($< 10^{-100}$, not shown here).}
\end{figure}

\begin{figure}
  \includegraphics{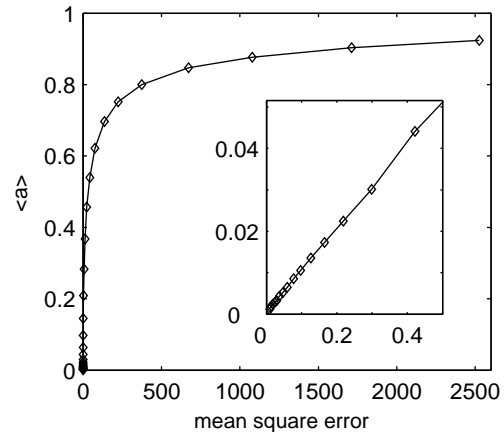}
  \caption{\label{fig:8}The average convergence measure $\la a\ra$ plotted
    against the corresponding mean square error $\la(\dfhat-\df)^2\ra$ of the
    two-sided free energy estimator. The inset shows an enlargement for small
    values of $\la a\ra$.}
\end{figure}

\begin{figure}
  \includegraphics{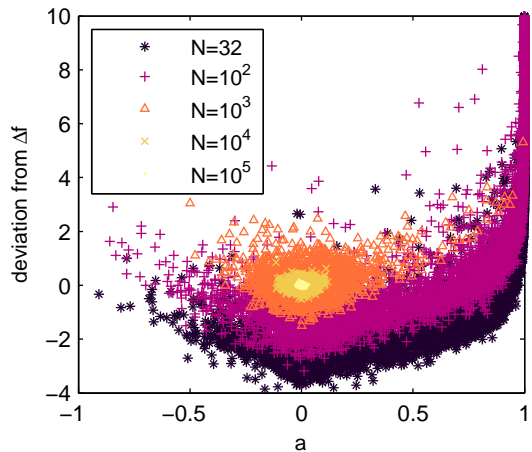}
  \caption{\label{fig:9}(Color online)
    A scatter plot of the deviation $\dfhat-\df$ versus the convergence measure
    $a$ for many individual estimates in dependence of the sample size $N$.
    Note that the majority of estimates belonging to $N=32$ and $N=100$ have
    large values of $\dfhat-\df$ well outside the displayed range with $a$
    being close to one.}
\end{figure}

\begin{figure}
  \includegraphics{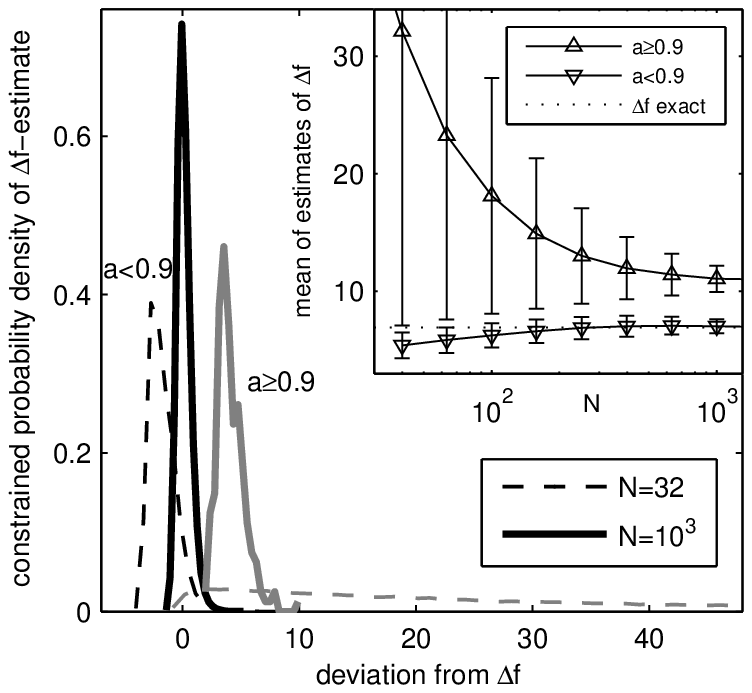}
  \caption{\label{fig:10}Estimated constrained probability densities
    $p(\dfhat|a\!<\!0.9)$ (black) and $p(\dfhat|a\!\ge\!0.9)$ (grayscale) for
    two different sample sizes $N$, plotted versus the deviation $\dfhat-\df$.
    The inset shows averaged estimates of $\df$ over the total sample size $N$
    subject to the constraints $a\ge0.9$ and $a<0.9$, respectively.}
\end{figure}

Choosing $\mu_0=1000$ and $\al=\frac{1}{2}$, i.e.\ $n_0=n_1$, we calculate free
energy estimates $\dfhat$ according to Eq.~\gl{benest} together with the
corresponding values of $a$ according to Eq.~\gl{a} for different total sample
sizes $N=n_0+n_1$. An example of a single running estimate and the
corresponding values of the convergence measure are depicted in
Fig.~\ref{fig:1}. Ten-thousand repetitions for each value of $N$ yield the
results presented in Figs.~\ref{fig:5}--\ref{fig:10}. To begin with, the top
panel of Fig.~\ref{fig:5} shows the averaged free energy estimates in
dependence of $N$, where the errorbars show $\pm$ the estimated square root of
the variance $\la(\dfhat-\langle\dfhat\rangle)^2\ra$. For small $N$, the bias
$\la\dfhat-\df\ra$ of free energy estimates is large, but becomes negligible
compared to the standard deviation for $N\gtrsim 5000$. This is a prerequisite
of the large $N$ limit, therefore we will view $N \approx 5000$ as the onset of
the large $N$ limit.

The bottom panel of Fig.~\ref{fig:5} shows the averaged values of the
convergence measure $a$ corresponding to the free energy estimates of the top
panel. Again, the errorbars are $\pm$ one standard deviation
$\sqrt{\la a^2\ra -\la a\ra^2}$, except that the upper limit is truncated for
small $N$, as $a<1$ holds. The trend of the averaged convergence measure
$\la a \ra$ is in full agreement with the general considerations given in the
previous section: for small $N$, $\la a \ra$ starts close to its upper bound,
decreases monotonically with increasing sample size, and converges towards zero
in the large $N$ limit. At the same time, its standard deviation converges to
zero, too. This indicates that single values of $a$ corresponding to single
estimates $\dfhat$ will typically be found close to zero in the large $N$
regime.

Noting that $a$ is defined as relative difference of the overlap estimators
$\Uhat_\al$ and $\Uhatvar_\al$ of first and second order, respectively, we can
understand the trend of the average convergence measure by taking into
consideration the average values $\la\Uhat_\al\ra$ and $\la\Uhatvar_\al\ra$,
which are shown in Fig.~\ref{fig:6}. For small sample sizes, $\Ua$ is
\textit{typically} underestimated by both, $\Uhat_\al$ and $\Uhatvar_\al$, with
$\Uhatvar_\al<\Uhat_\al$.

The convergence measure takes advantage of the different convergence times of
the overlap estimators: $\Uhatvar_\al$ converges somewhat slower than
$\Uhat_\al$, ensuring that $a$ approaches zero right after $\dfhat$ has
converged. The large standard deviations shown as errorbars in Fig.~\ref{fig:6}
do not carry over to the standard deviation of $a$, because $\Uhat_\al$ and
$\Uhatvar_\al$ are strongly correlated, as is impressively visible in
Fig.~\ref{fig:7}. The estimated correlation coefficient
\begin{align}
  \frac{ \la \big(\Uhatvar_\al-\langle\Uhatvar_\al\rangle \big)
    \big(\Uhat_\al-\langle\Uhat_\al\rangle\big) \ra }{
    \sqrt{\Var\big(\Uhatvar_\al\big) \Var\big(\Uhat_\al\big)} }
\end{align}
is about $0.97$ (!) for the entire range of sample sizes $N$. In good
approximation, $\Uhat_\al$ and $\Uhatvar_\al$ are related to each other
according to a power law, $\Uhatvar_\al \approx c_N \Uhat_\al^{\ \gamma_N}$,
where the exponent $\gamma_N$ and the prefactor $c_N$ depend on the sample size
$N$ (and $\al$). We note that $\gamma_N$ has a phase-transition-like behavior:
for small $N$, it stays approximately constant near two; right before the onset
of the large $N$ limit, it shows a sudden switch to a value close to one where
it finally remains.

Figure \ref{fig:8} accents the decrease of the average $\la a\ra$ with
decreasing mean square error \gl{msefull} of two-sided estimation. The small
$N$ behavior is given by the upper right part of the graph, where $\la a\ra$ is
close to its upper bound together with a large mean square error of $\dfhat$.
With increasing sample size, the mean square error starts to drop somewhat
sooner than $\la a\ra$, however, at the onset of the large $N$ limit, they drop
both and suggest a linear relation, as can be seen in the inset for small
values of $\la a\ra$. The latter shows that $\la a \ra$ decreases to zero
proportional to $\frac{1}{N}$ for large $N$ (this is confirmed by a direct
check, but not shown here).

The next point is to clarify the correlation of single values of the
convergence measure with their corresponding free energy estimates. For this
issue, figure \ref{fig:9} is most informative, showing the deviations
$\dfhat-\df$ in dependence of the corresponding values of $a$ for many
individual observations. The figure makes clear that there is a \textit{strong
  relation, but no one-to-one correspondence} between $a$ and $\dfhat-\df$: For
large $N$, both $a$ and $\dfhat-\df$ approach zero with very weak correlations
between them. However, the situation is different for small sample sizes $N$
where the bias $\la\dfhat-\df\ra$ is considerably large. There, the typically
observed large deviations occur together with values of $a$ close to the upper
bound, whereas the atypical events with small (negative) deviations come
together with values of $a$ well below the upper limit. Therefore, small values
of $a$ detect exceptional events if $N$ is well below the large $N$ limit, and
ordinary events if $N$ is large.

To make this relation more visible, we split the estimates $\dfhat$ into the
mutually exclusive events $a\geq 0.9$ and $a< 0.9$. The statistics of the
$\dfhat$ values within these cases are depicted in the inset of
Fig.~\ref{fig:10}, where normalized histograms, i.e.\ estimates of the
constrained probability densities $p(\dfhat|a\!\geq\!0.9)$ and
$p(\dfhat|a\!<\!0.9)$ are shown. The unconstrained probability density of
$\dfhat$ can be reconstructed from a likelihood weighted sum of the constrained
densities,
$p(\dfhat) = p(\dfhat|a\!\geq\!0.9) p_{\scriptscriptstyle a\geq0.9}
+ p(\dfhat|a\!<\!0.9) p_{\scriptscriptstyle a<0.9}$. The likelihood ratios read
$p_{\scriptscriptstyle a\geq0.9}/p_{\scriptscriptstyle a<0.9} = 6.2$ and $0.002$
for $N=32$ and $1000$, respectively. Finally, the inset of Fig.~\ref{fig:10}
shows the average values of constrained estimates $\dfhat$ over $N$ with
errorbars of $\pm$ one standard deviation, in dependence of the condition on
$a$.

\subsection{Gaussian work densities}

For the second example the work-densities are chosen to be Gaussian,
\begin{align}
  p_i(w) = \frac{1}{\sigma\sqrt{2\pi}}
  e^{-\frac{(w-\mu_i)^2}{2\sigma^2}}, \quad w\in\mathds{R},
\end{align}
$i=0,1$. The fluctuation theorem \gl{fth} demands both densities to have the
same variance $\sigma^2$ with mean values $\mu_0=\df+\frac{1}{2}\sigma^2$ and
$\mu_1=\df-\frac{1}{2}\sigma^2$. Hence, $\pf$ and $\pr$ are symmetric to each
other with respect to $\df$, $\pf(\df+w)=\pr(\df-w)$. As a consequence of this
symmetry, the two-sided estimator with equal sample sizes $n_0$ and $n_1$,
i.e.\ $\al=0.5$, is unbiased for any $N$. However, this does not mean that the
limit of large $N$ is reached immediately.

In analogy to the previous example, we proceed in presenting the statistical
properties of $a$. Choosing $\sigma = 6$ and without loss of generality
$\df=0$, we carry out $10^4$ estimations of $\df$ over a range of sample sizes
$N$. The forward fraction is chosen to be equal to $\al=0.5$, and for
comparison, $\al=0.999$, and $\al=0.99999$, respectively. In the latter two
cases, the two-sided estimator is biased for small $N$. We note that $\al=0.5$
is always the optimal choice for symmetric work densities which minimizes the
asymptotic mean square error \gl{mse} with respect to $\al$ \cite{Hahn2009b}.

\begin{figure}
  \includegraphics{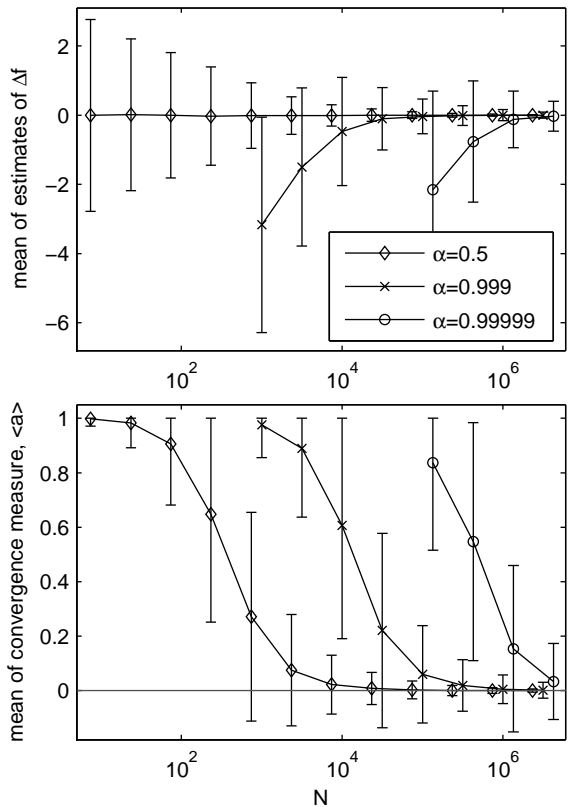}
  \caption{\label{fig:11}Gaussian work densities result in the displayed
    averaged estimates of $\df$. For comparison, three different fractions
    $\al$ of forward work values are used (top). Average values of the
    convergence measure $a$ corresponding to the estimates of the top panel
    (bottom).}
\end{figure}

Comparing the top and the bottom panel of Fig.~\ref{fig:11}, which show the
statistics (mean value and standard deviation as errorbars) of the observed
estimates $\dfhat$ and of the corresponding values of $a$, we find a coherent
behavior for all three cases of $\al$ values. The trend of the average
$\la a\ra$ shows in all cases the same features in agreement with the trend
found for exponential work densities. 

\begin{figure}
  \includegraphics{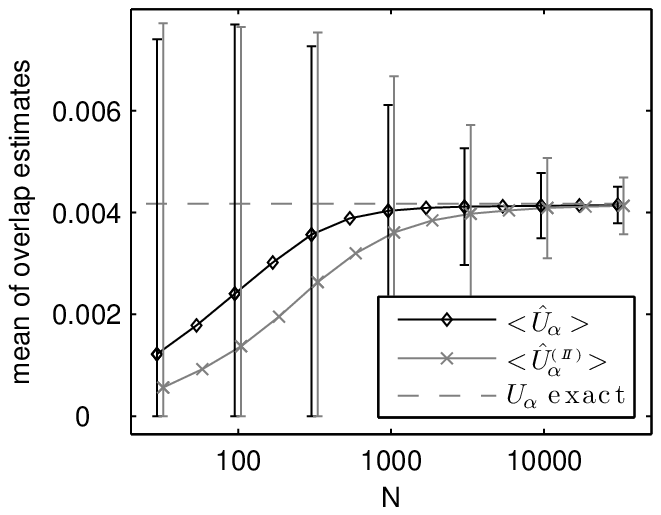}
  \caption{\label{fig:12}Mean values of overlap estimates $\Uhat_\al$ and
    $\Uhatvar_\al$ of first and second order ($\al=0.5$).}
\end{figure}

As before, the characteristics of $a$ are understood by the slower convergence
of $\Uhatvar_\al$ compared to that of $\Uhat_\al$, as can be seen in
Fig.~\ref{fig:12}. A scatter plot of $\Uhatvar_\al$ versus $\Uhat_\al$ looks
qualitatively like Fig.~\ref{fig:7}, but is not shown here.

\begin{figure}
  \includegraphics{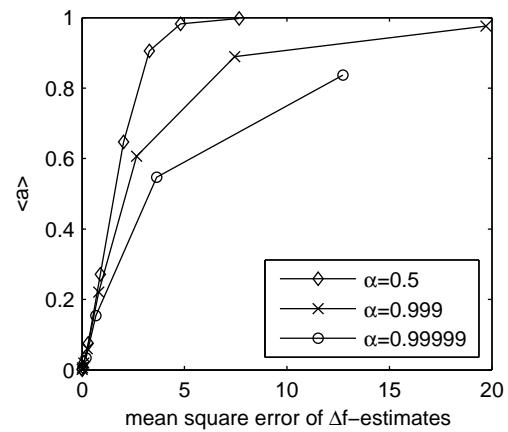}
  \caption{\label{fig:13}The average convergence measure $\la a\ra$ plotted
    against the corresponding mean square error $\la(\dfhat-\df)^2\ra$ of free
    energy estimates in dependence of $N$.}
\end{figure}

Figure \ref{fig:13} compares the average convergence measures as functions of
the mean square error of $\dfhat$ for the three values of $\al$. For the range
of small $\la a\ra$, all three curves agree and are linear. Again $\la a \ra$
decreases proportionally to $\frac{1}{N}$ for large $N$. Noticeable for small
$N$ is the shift of $\la a \ra$ towards smaller values with increasing $\al$.
This results from the definition of $a$: the upper bound $1-\Uhat_\al$ of $a$
tends to zero in the limits $\al\to 0,1$, as then $\Uhat_\al\to1$.

\begin{figure}
  \includegraphics{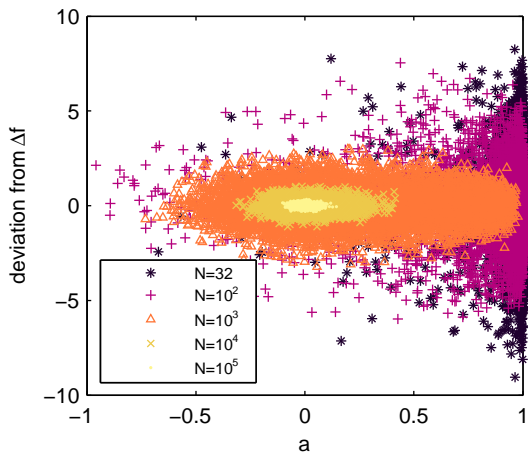}
  \caption{\label{fig:14}(Color online) A scatter plot of the deviation
    $\dfhat-\df$ versus the convergence measure $a$ for many individual
    estimates in dependence of the sample size $N$ ($\al=0.5$).}
\end{figure}

The relation of single free energy estimates $\dfhat$ with the corresponding
$a$ values can be seen in the scatter plot of Fig.~\ref{fig:14}. The mirror
symmetry of the plot originates from the symmetry of the work densities and the
choice $\al=0.5$, i.e.\ of the unbiasedness of the two-sided estimator. Opposed
to the foregoing example, the correlation between $\dfhat-\df$ and $a$ vanishes
for any value of $N$. Despite the lack of any correlation, the figure reveals a
strong relation between the deviation $\dfhat-\df$ and the value of $a$: they
converge equally to zero for large $N$.

\begin{figure}
  \includegraphics{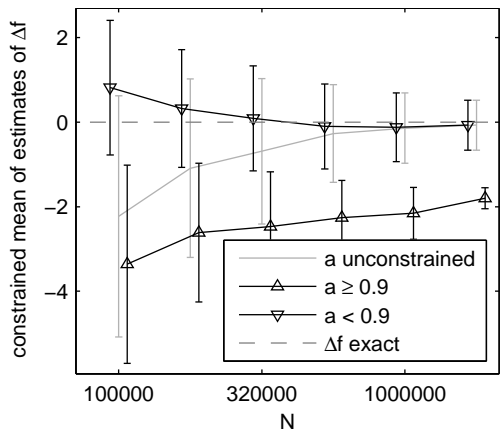}
  \caption{\label{fig:15}Averaged two-sided estimates of $\df$ in dependence of
    the total sample size $N$ for the constraints $a\ge0.9$, $a$ unconstrained,
    and $a<0.9$ ($\al=0.99999$).}
\end{figure}

Last, figure \ref{fig:15} shows averages of constrained $\df$ estimates for the
mutually exclusive conditions $a\geq0.9$ and $a<0.9$, now with $\al=0.99999$ in
order to incorporate some bias. We observe the same characteristics as before,
cf.\ the inset of Fig.~\ref{fig:10}: The condition $a<0.9$ filters the
estimates $\dfhat$ which are closer to the true value.

\subsection{The general case}

The characteristics of the convergence measure are dominated by contributions
of work densities inside and near the region where the overlap density
$\pa(w)$, Eq.~\gl{pa}, has most of its mass. We call this region the overlap
region. In the overlap region, the work densities may have one of the following
characteristic relation of shape:
\begin{enumerate}
\item Having their maxima at larger and smaller values of work, respectively,
  the forward and reverse work densities both drop towards the overlap region.
  Hence, any of both densities sample the overlap region by rare events, only,
  which are responsible for the behavior of the convergence measure.
\item Both densities decrease with increasing $w$ and the overlap region is
  well sampled by the forward work density compared with the reverse density.
  Especially the ``rare'' events $w<\df$ of forward direction are much more
  available than the rare events $w>\df$ of reverse direction. Hence, more or
  less typical events of one direction together with atypical events of the
  other direction are responsible for the behavior of the convergence measure.
  Likewise if both densities increase with $w$.
\item More generally, the work densities are some kind of interpolation between
  the above two cases.
\item Finally, there remain some exceptional cases. For instance, if the
  forward and reverse work densities have different support or if they do not
  obey the fluctuation theorem at all.
\end{enumerate}
With respect to the exceptional case, the convergence measure fails to work,
since it requires that the forward and reverse work densities have the same
support and that the densities are related to each other via the fluctuation
theorem \gl{fth}.

In all other cases, the convergence measure certainly will work and will show a
similar behavior, regardless of the detailed nature of the densities. This can
be explained as follows. In the preceding subsections, we have investigated
exponential and Gaussian work densities, two examples that differ in their very
nature. While exponential work densities cover case number two, and Gaussians
cover case number one, they show the same characteristics of $a$. This means
that the characteristics of the convergence measure are insensitive to the
individual nature of the work densities as long as they have the same support
and obey the fluctuation theorem.

To this end, we want to point to some subtleties in the text of the actual
paper. While the measure of convergence is robust with respect to the nature of
work densities, some heuristic or pedagogic explanations in the text are
written with regard to the typical case number one, where the overlap region is
sampled by rare events, only. This concerns mainly Sec.~\ref{sec:2} where we
speak about effective cut-off values in the context of the neglected tail
model. These effective cut-off values would become void if we would try to
explain the bias of exponential work densities qualitatively via the neglected
tail model. Also the explanations in the text of the next section are mainly
focused on the typical case number one. This concerns the passages where we speak
about rare events. Nevertheless, the main and essential statements are valid
for all cases.

The most important property of $a$ is its almost \textit{simultaneous}
convergence with the free energy estimator $\dfhat$ to an \textit{\`a priori
  known} value. This fact is used to develop a convergence criterion in the
next section.

\section{The convergence criterion}\label{sec:5}

Elaborated the statistical properties of the convergence measure, we are
finally interested in the convergence of a \textit{single} free energy
estimate. In contrast to averages of many independent running estimates,
estimates based on individual realization are not smooth in $N$, see
e.g.\ Fig.~\ref{fig:1}.

For small $N$, typically $\Uhatvar_\al$ underestimates $\Ua$ more than
$\Uhat_\al$ does, pushing $a$ close to its upper bound. With increasing $N$,
$\dfhat$ starts to ``converge''; typically in a non-smooth manner. The
convergence of $\dfhat$ is triggered by the occurrence of rare events. Whenever
such a rare event in the important tails of the work densities gets sampled,
$\dfhat$ jumps, and between such jumps, $\dfhat$ stays rather on a stable
plateau. The measure $a$ is triggered by the same rare events, but the changes
in $a$ are smaller, unless convergence starts happening. Typically, the rare
events that bring $\dfhat$ near to its true value are the rare events which
change the value of $a$ drastically. In the typical case, these rare events let
$a$ even undershoot below zero, before $\dfhat$ and $a$ finally converge.

The features of the convergence measure,
\begin{enumerate}
\item it is bounded, $a\in(-1,\,1-\Uhat_\al]$,
\item it starts for small $N$ at its upper bound,
\item it converges to a known value, $a\to0$,
\item and typically it converges almost simultaneously with $\dfhat$,
\end{enumerate}
simplify the task of monitoring the convergence significantly, since it is far
easier to compare estimates of $a$ with the known value zero than the task of
monitoring convergence of $\dfhat$ to an unknown target value. The
characteristics of the convergence measure enable us to state: typically, if a
is close to zero, $\dfhat$ has converged.

Deviations from the typical situation are possible. For instance, $\dfhat$ may
not show such clear jumps, neither may $a$. Occasionally, $\dfhat$ and $a$, may
also fluctuate exceedingly strong. Thus, a single value of $a$ close to zero
does not guarantee convergence of the free energy estimate as can be seen from
some few individual events in the scatter plot of Fig.~\ref{fig:14} that fail a
correct estimate while $a$ is close to zero. A single random realization may
give rise to a fluctuation that brings $a$ close to zero by chance, a fact that
needs to be distinguished from $a$ having converged to zero. The difference
between random chance and convergence is revealed by increasing the sample
size, since it is highly unlikely that $a$ stays close to zero by random. It is
the \textit{behavior} of $a$ with increasing $N$, that needs to be taken into
account in order to establish an equivalence between $a\to0$ and $\dfhat\to\df$.

This allows us to state the convergence criterion:
\begin{itemize}
\item[] \textit{if $a$ fluctuates close around zero, convergence is assured},
\end{itemize}
implying that if $a$ fluctuates around zero, $\dfhat$ fluctuates around its
true value $\df$, the bias vanishes, and the mean square error reaches its
asymptotics which can be estimated using Eq.~\gl{Xhat}. $a$ fluctuating close
around zero means that it does so over a suitable range of sample sizes, which
extends over an order of magnitude or more.

\section{Application}\label{sec:6}

As an example, we apply the convergence criterion to the calculation of
the excess chemical potential $\mu^{ex}$ of a Lennard Jones fluid. Using
Metropolis Monte Carlo simulation \cite{Metropolis1953} of a fluid of $N_p$
particles, the forward work is defined as energy increase when inserting at
random a particle into a given configuration \cite{Widom1963}, whereas the
reverse work is defined as energy decrease when a random particle is deleted
from a given $N_p+1$-particle configuration. The densities $p_0(w)$ and
$p_1(w)$ of forward and reverse work obey the fluctuation theorem \gl{fth} with
$\df=\mu^{ex}$ \cite{Hahn2009a}. Thus, Bennett's acceptance ratio method can be
applied to the calculation of the chemical potential.

Details of the simulation are reported in Ref.~\cite{Hahn2009a}. Here, the
parameter values chosen read: $N_p=120$, reduced Temperature $T^*=1.2$, and
reduced density $\rho^*=0.5$.

\begin{figure}
  \includegraphics{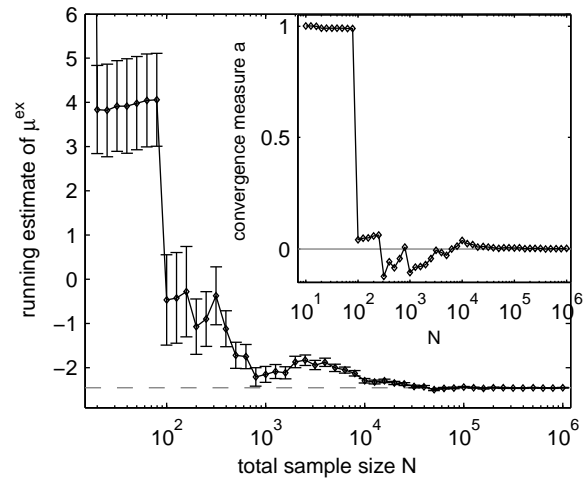}
  \caption{\label{fig:16} Running estimates of the excess chemical potential
    $\mu^{ex}$ in dependence of the sample size $N$ ($\al=0.9$). The inset
    displays the corresponding values of the convergence measure $a$.}
\end{figure}

Drawing work values up to a total sample size of $10^6$ with fraction $\al=0.9$
of forward draws (which will be a near-optimal choice \cite{Hahn2009b}), the
successive estimates of the chemical potential together with the corresponding
values of the convergence measure are shown in Fig.~\ref{fig:16}. The dashed
horizontal line does not show the exact value of $\mu^{ex}$, which is unknown,
but rather the value of the last estimate with $N=10^6$. Taking a closer look
on the behavior of the convergence measure with increasing $N$, we observe $a$
near unity for $N\leq 10^2$, indicating the low $N$ regime and the lack of
observing rare events. Then, a sudden drop near to zero happens at
$N=10^2$, which coincides with a large jump of the estimate of $\mu^{ex}$,
followed by large fluctuations of $a$ with strong negative values in the regime
$N=10^2$ to $10^4$. This behavior indicates that the important but rare events
which trigger the convergence of the $\mu^{ex}$ estimate are now sampled, but
with strongly fluctuating relative frequency, which in specific cases causes the negative
values of $a$ (because of too many rare events!). Finally, with
$N > 10^4$, $a$ equilibrates and converges to zero. The latter is observed over
two orders of magnitude, such that we can conclude that the latest estimate of
$\mu^{ex}$ with $N=10^6$ has surely converged and yields a reliable value of
the chemical potential. The confidence interval of the estimate can safely be
calculated as the square root of Eq.~\gl{mse} (one standard deviation), and we
obtain explicitly $\widehat{\mu^{ex}} = -2.451 \pm 0.005$.

\begin{figure}
  \includegraphics{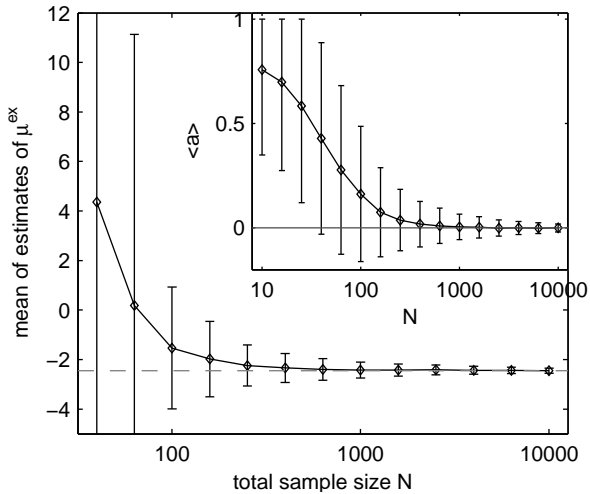}
  \caption{\label{fig:17} Statistics of estimates of the excess chemical
    potential: shown are the average value and the standard deviation (as
    errorbars) in dependence of the sample size $N$. The statistics of the
    corresponding values of the convergence measure is shown in the inset.}
\end{figure}

Interested in the statistical behavior of $a$ for the present application, we
carried out 270 simulation runs up to $N=10^4$ to obtain the average values and
standard deviations of $\widehat{\mu^{ex}}$ and $a$ which are depicted in
Fig.~\ref{fig:17}. The dashed line marks the same value as that in
Fig.~\ref{fig:16}. Again, we observe the same qualitative behavior of $a$ as in
the foregoing examples of Sec.~\ref{sec:5}, especially a positive average value
of $\la a \ra$ and a convergence to zero which occurs simultaneously with the
convergence of Bennett's acceptance ratio method.

\section{Conclusions}\label{sec:7}

Since its formulation a decade ago, the Jarzynski equation and the Crooks
fluctuation theorem gave rise to enforced research of nonequilibrium techniques
for free energy calculations. Despite the variety of new methods, in general
little is known about their statistical properties. In particular, it is often
unclear whether the methods actually converge to the desired value of the free
energy difference $\df$, and if so, it remains in question whether convergence
happened within a given calculation. This is of great concern, as usually the
calculations are strongly biased before convergence starts happening. In
consequence, it is impossible to state the result of a single calculation of
$\df$ with a reliable confidence interval unless a convergence measure is
evaluated.

In this paper, we presented and tested a quantitative measure of convergence
for two-sided free energy estimation, i.e.\ Bennett's acceptance ratio method,
which is intimately related to the fluctuation theorem. From this follows a
criterion for convergence relying on monitoring the convergence measure $a$
within a running estimation of $\df$. The heart of the convergence criterion is
the nearly simultaneous convergence of the free energy calculation and the
convergence measure $a$. Whereas the former converges towards the unknown value
$\df$, which makes it difficult or even impossible to decide when convergence
actually takes place, the latter converges to an \textit{\`a priori known}
value. If convergence is detected with the convergence criterion, the
calculation results in a reliable estimate of the free energy difference
together with a precise confidence interval.
 
\appendix

\section{}\label{appendix.A}

The derivation of inequality \gl{mseineq} relies on the close connection
between the overlap $\Ua$ and the overlap area $\mathcal{A}$,
\begin{widetext}
  \begin{align}
    & \Ua = \int \frac{\pf\pr}{\al\pf+\be\pr} dw
    \ge \int \frac{\pf\pr}{(\al+\be)\max\{\pf,\pr\}} dw
    = \int \min\{\pf,\pr\} dw = \mathcal{A}, \\
    & U_{\frac{1}{2}} = 2 \int \frac{1}{1/\pr + 1/\pf} dw
    < 2 \int \min\{\pf,\pr\} dw = 2\mathcal{A}.
  \end{align}
  Together with the inequality $\frac{1}{2}X(N,\frac{1}{2})\le X(N,\al)$ of
  Bennett \cite{Bennett1976}, we obtain
  \begin{align}
    \frac{1-2\mathcal{A}}{N\mathcal{A}}
    < \frac{1-U_{\frac{1}{2}}}{\frac{1}{2}NU_{\frac{1}{2}}}
    = \frac{1}{2}X(N,\frac{1}{2}) \le X(N,\al)
    \le \frac{1}{N}\frac{1}{\al\be}\big(\frac{1}{\mathcal{A}}-1\big)
  \end{align}
  which directly yields inequality \gl{mseineq}.

  \section{}\label{appendix.B}

  Inequality \gl{Uineq} can be obtained as follows. Noting that
  $t_c(w)< \frac{1}{\al}$ and $b_c(w)< \frac{1}{\be}$, cf. Eqs.~\gl{fermidef} we have
  \begin{align}
    & 2\Uhat_\al
    = \widebar{t_{\scriptscriptstyle \dfhat}}^{\scriptscriptstyle (1)}
    + \widebar{b_{\scriptscriptstyle \dfhat}}^{\scriptscriptstyle (0)}
    > \al\widebar{t_{\scriptscriptstyle \dfhat}^2}^{\scriptscriptstyle (1)}
    + \be\widebar{b_{\scriptscriptstyle \dfhat}^2}^{\scriptscriptstyle (0)}
    = \Uhatvar_\al \qquad \qquad \text{and further} \\
    & \Uhatvar_\al
    =  \Uhat_\al^2
    + \al \widebar{(t_{\scriptscriptstyle \dfhat}
      -\Uhat_\al)^2}^{\scriptscriptstyle (1)}
    + \be \widebar{(b_{\scriptscriptstyle \dfhat}
      -\Uhat_\al)^2}^{\scriptscriptstyle (0)}
    \ge \Uhat_\al^2,
  \end{align}
  which results in Eq.~\gl{Uineq}.
\end{widetext}

\section{}\label{appendix.C}

The errorbars in Figs.~\ref{fig:1} and \ref{fig:16} are obtained via the
error-propagation formula for the variance of Bennett's acceptance ratio
method.

A possible estimate $\widehat{\sigma}_{ep}^2$ of the variance of the two-sided free
energy estimator obtained from error-propagation reads
\begin{align}
 \widehat{\sigma}_{ep}^2 = \frac{1}{n_1}
 \frac{\widebar{t_{\scriptscriptstyle \dfhat}^2}^{\scriptscriptstyle (1)}
- {\widebar{t_{\scriptscriptstyle \dfhat}}^{\scriptscriptstyle (1)}}^2}
      {{\widebar{t_{\scriptscriptstyle \dfhat}}^{\scriptscriptstyle (1)}}^2}
      + \frac{1}{n_0}
      \frac{\widebar{b_{\scriptscriptstyle \dfhat}^2}^{\scriptscriptstyle (0)}
        - {\widebar{b_{\scriptscriptstyle \dfhat}}^{\scriptscriptstyle (0)}}^2}
           {{\widebar{b_{\scriptscriptstyle \dfhat}}^{\scriptscriptstyle (0)}}^2}
\end{align}
Alternatively, $\widehat{\sigma}_{ep}^2$ can be expressed through the overlap estimates
$\Uhat_\al$ and $\Uhatvar_\al$ of first and second order, Eqs.~\gl{Uhatnew} and
\gl{Uhat2},
\begin{align}
 \widehat{\sigma}_{ep}^2
 = \frac{1}{\al\be N} \frac{\Uhatvar_\al - \Uhat_\al^2}{\Uhat_\al^2}.
\end{align}
In the limit of large $N$, $\widehat{\sigma}_{ep}^2$ converges to the asymptotic mean
square error $X(N,\al)$, Eq.~\gl{mse}. An upper bound on $\widehat{\sigma}_{ep}^2$
follows from inequality \gl{Uineq2}:
\begin{align}
  \widehat{\sigma}_{ep}^2 \leq 2 - \frac{1}{\al\be N}.
\end{align}
Finally let us mention that the convergence measure $a$, Eq.~\gl{a}, is closely
related to the relative difference of the estimated asymptotic mean square
error $\hat{X}$, Eq.~\gl{Xhat}, and $\widehat{\sigma}_{ep}^2$:
\begin{align}
  a = (1-\Uhat_\al) \frac{\hat{X} - \widehat{\sigma}_{ep}^2}{\hat{X}}.
\end{align}

\section{}\label{appendix.D}
Consider the family $\widehat{\phi}(c)$ of $\df$ estimators, parametrized by
the real number $c$ \cite{Bennett1976}:
\begin{align}
 \widehat{\phi}(c) = c + \ln\frac{\widebar{t_{c}}^{\scriptscriptstyle (1)}}{
   \widebar{b_{c}}^{\scriptscriptstyle (0)}}.
\end{align}
For any fixed value of $c$, $\widehat{\phi}(c)$ defines a consistent estimator
of $\df$, $\widehat{\phi}(c)\overset{N\to\infty}{\longrightarrow}\df \quad
\forall c$. For finite $N$, however, the performance of the estimator strongly
depends on $c$. The (optimal) two-sided estimate \gl{benest} is obtained by the
additional condition $\widehat{\phi}(c)=c$, such that
$\widebar{t_{c}}^{\scriptscriptstyle (1)}=\widebar{b_{c}}^{\scriptscriptstyle (0)}$
holds, and thus $c=\dfhat$. A possible measure for the sensibility of the
estimate $\widehat{\phi}(c)$ on $c$ is it's derivative with respect to $c$.
Using $\frac{\partial}{\partial c}t_c = -\beta t_c b_c$,
$\frac{\partial}{\partial c}b_c = \alpha t_c b_c$, and $\al t_c +\be b_c =1$,
we obtain
\begin{align}
  \frac{\partial}{\partial c}\widehat{\phi}(c)
  = -1 + \al \frac{\widebar{t_{c}^2}^{\scriptscriptstyle (1)}}{
    \widebar{t_{c}}^{\scriptscriptstyle (1)}}
  + \be \frac{\widebar{b_{c}^2}^{\scriptscriptstyle (0)}}{
    \widebar{b_{c}}^{\scriptscriptstyle (0)}}.
\end{align}
Taking the derivative at $c=\dfhat$ directly results in the convergence measure
$a$,
\begin{align}
  \frac{\partial}{\partial c}\widehat{\phi}(c)|_{\dfhat} = -a.
\end{align}

\end{document}